\documentclass[cits]{PoS-pdf}
\usepackage{epsfig}
\usepackage{cite}
\usepackage{amsmath}
\usepackage{eufrak}
\usepackage{amssymb}
\usepackage{eufrak}
\usepackage{color}
\usepackage{colordvi}
\usepackage{float}

\def\TCop{\textsuperscript{\textcopyright}}

\definecolor{darkgreen}      {rgb}{0.0,0.6,0.0}
\let\normalcolor\relax 

\newcommand{\ep}{\varepsilon}
\newcommand{\N}{\nonumber}
\newcommand{\GeV}{\rm GeV}

\title{\footnotesize
{DESY 12-031, DO-TH-12/07, SFB/CPP-12-20, LPN 12--047  \hfill {\tt arxiv:1205.4991}
\vspace*{5mm}\noindent
\\
\LARGE Why Precision?}}

\ShortTitle{Why Precision?}

\author{\speaker{Johannes~Bl\"umlein}\\
Deutsches Elektronen-Synchrotron, DESY, Zeuthen, Platanenalle 6, D-15735 Zeuthen, Germany.}

\abstract{
Precision measurements together with exact theoretical calculations have led to steady
progress in fundamental physics. A brief survey is given on  recent developments and
current achievements in the field of perturbative precision calculations in the Standard Model
of the Elementary Particles and their application in current high energy collider data analyses.}

\FullConference{
10th International Symposium on Radiative Corrections (Applications of Quantum Field Theory to Phenomenology), RADCOR
2011
Mamallapuram, India
September 26-30, 2011}

\begin{document}
\section{Introduction}

\noindent
Precision matters. Any progress in the exact sciences relies both on precise
measurements and highly accurate theoretical calculations. Many of the fundamental 
laws of physics had unavoidably to be found whenever precise data were described by 
theoretical concepts, often within a new framework of relations. The Rudolphine 
Tables of the late Tycho Brahe \cite{H1} led J.~Kepler to derive his laws \cite{H2} 
and later I.~Newton the law of gravity \cite{H3}. A.~Michelson's experiments \cite{H4} 
led A.~Einstein to Special Relativity \cite{H5}, with numerous experimental confirmations 
in flat space-time.~\footnote{At the start of this conference Ref.~\cite{OPERA} appeared, which is 
currently in course of re-analysis \cite{OPERA1}. The result has not been confirmed by a
more recent measurement \cite{Antonello:2012hg}.} The 
accurate measurement of the 
black-body radiation by F.~Kurlbaum, H.~Rubens, O.~Lummer and E.~Pringsheim \cite{H6} 
enforced M.~Planck to quantize the action \cite{H7}. The term measurement of the 
spectral series by J.~Balmer and others \cite{H8} led N.~Bohr \cite{H9} to construct 
his model of the hydrogen atom. O.~Frisch and O.~Stern discovered the anomalous 
magnetic moment of the proton \cite{H10}. About 35 years later this phenomenon
could be explained by finding the short-distance structure of nucleons as quark 
and gluon partons by the MIT-SLAC experiments \cite{H11}. The discovery of the
weak neutral currents by Gargamelle and the polarization-symmetry in deep-inelastic
scattering in brilliant experiments clearly indicated the existence of 
the $W^\pm$ and $Z^0$ bosons \cite{H12}, which were discovered by UA1 and UA2 \cite{H13} 
shortly 
after. M. Veltman \cite{H14} found the quadratic mass effects of fermions in the 
1-loop electro-weak radiative corrections. The future inclusive electro-weak precision 
measurements allowed to locate the mass of the top quark, where it was found at
Tevatron later \cite{H15}. At present, similar more stringent constraints, exploiting the 
known QCD and electro-weak corrections to $e^+e^-$ and $pp$ resp. $\bar{p}$ scattering 
is setting tighter and tighter mass limits for the Higgs boson \cite{H16}.

Precision measurements together with precision calculations in the framework of the present 
Standard Model of the elementary particles and its possible renormalizable extensions allow 
to search for new phenomena. One may thus expect that the discoveries mentioned before will 
be followed by various more using precision methods. At the experimental side, key topics 
are 
the detailed exploration of the heavy quark sector ($b, t$) at $B$-factories, the LHC and a 
future ILC. The masses and mixing parameters in the neutrino sector have to be measured more 
precisely at $\nu$-facilities and using astrophysical observations. Another central question 
concerns the precision measurement of the coupling constants, in particular also of $\alpha_s(M_Z)$,
which is least known \cite{Bethke:2011tr}. A major task for the experiments at the LHC
consists in the search for the Higgs boson of the Standard Model and possible extensions.
If it turns out that the fundamental fermions and bosons do not acquire their masses 
through the Higgs mechanism, the interaction of the weak bosons will become strong at 
high energy scales. To investigate this potential phenomenon the LHC experiments 
need to measure the interaction of weak bosons very precisely. The final task in exploring
the new kinematic domain at the LHC is to search for new particles and forces. During this
conference two surveys on the present results and the physics potential of ATLAS \cite{KATZY} 
and CMS \cite{PIERI} were given.

On the side of theoretical computations the level of 4- and 5-loop massless and 
massive calculations in QCD for zero-scale quantities are performed. $2 \rightarrow n$ 
scattering processes within the electro-weak theory, QCD, and the MSSM are carried out up 
to the
2-loop level. Unpolarized and polarized QCD calculations reached the 3-loop level for single 
differential distributions. The cross sections for $ep \rightarrow 3$ jets and  $pp 
\rightarrow 2$ jets at 
next-to-next-to leading order (NNLO) are underway. The predictions for many processes 
are improved adding appropriate resummations. At the technology side, many calculational 
tools
are developed, partly in close collaboration with mathematical groups. In this way, both 
the 
calculations and numerical simulations were greatly improved. In case of various experimental 
measurements it is presently needed to understand QCD corrections at a level of better than 1\%, 
which requires calculations at the NNLO level and higher.  The precise understanding of all 
`backgrounds' to the anticipated discoveries is of essential importance and requires all 
the 
ongoing theoretical efforts. 
\normalsize
\begin{center}
\colorbox{red}{\textcolor{white}{Quantum Field Theory}}

\vspace*{5mm}
{\LARGE $\Downarrow$}

\vspace*{0.5cm}
\colorbox{darkgreen}{\textcolor{yellow}{Scattering Cross Sections}}
      
\vspace*{0.5cm}
{\LARGE $\Uparrow$}
\hspace*{2cm}
{\LARGE $\Uparrow$}

\vspace*{0.5cm}
\colorbox{blue}{\textcolor{yellow}{Mathematics}}  \hspace*{3cm} 
\colorbox{yellow}{\textcolor{blue}{Algorithmics}}
\end{center}
{The precision measurements~at 
the large scale high-energy} 
facilities like HERA, Tevatron, LHC, the $B$-factories, 
precision measurements in $\nu$-physics and planned facilities like the EIC \cite{EIC} and ILC 
\cite{ILC} have driven the theoretical calculations to a much higher level. Many contemporary
quantum field theoretic calculations are performed referring to new mathematical methods and
are based on an intense use of computer algebra and combinatoric algorithms to end up with
precision predictions for scattering cross sections. In this way quantum field theories are 
understood on the perturbative level a lot better.  

\restylefloat{figure}
\begin{figure}[H]
\begin{center}
\includegraphics[scale=0.22,angle=0]{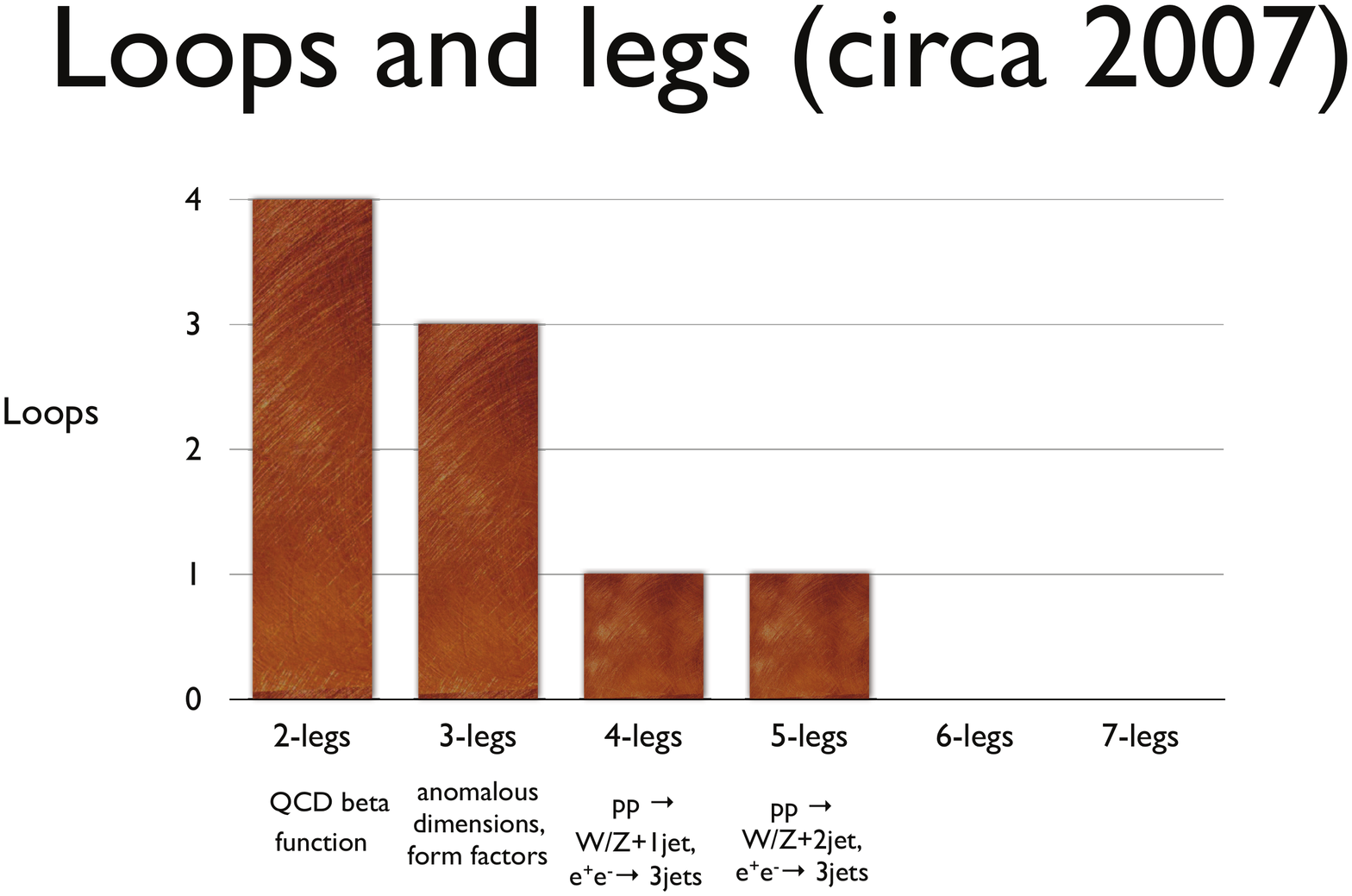}
\includegraphics[scale=0.22,angle=0]{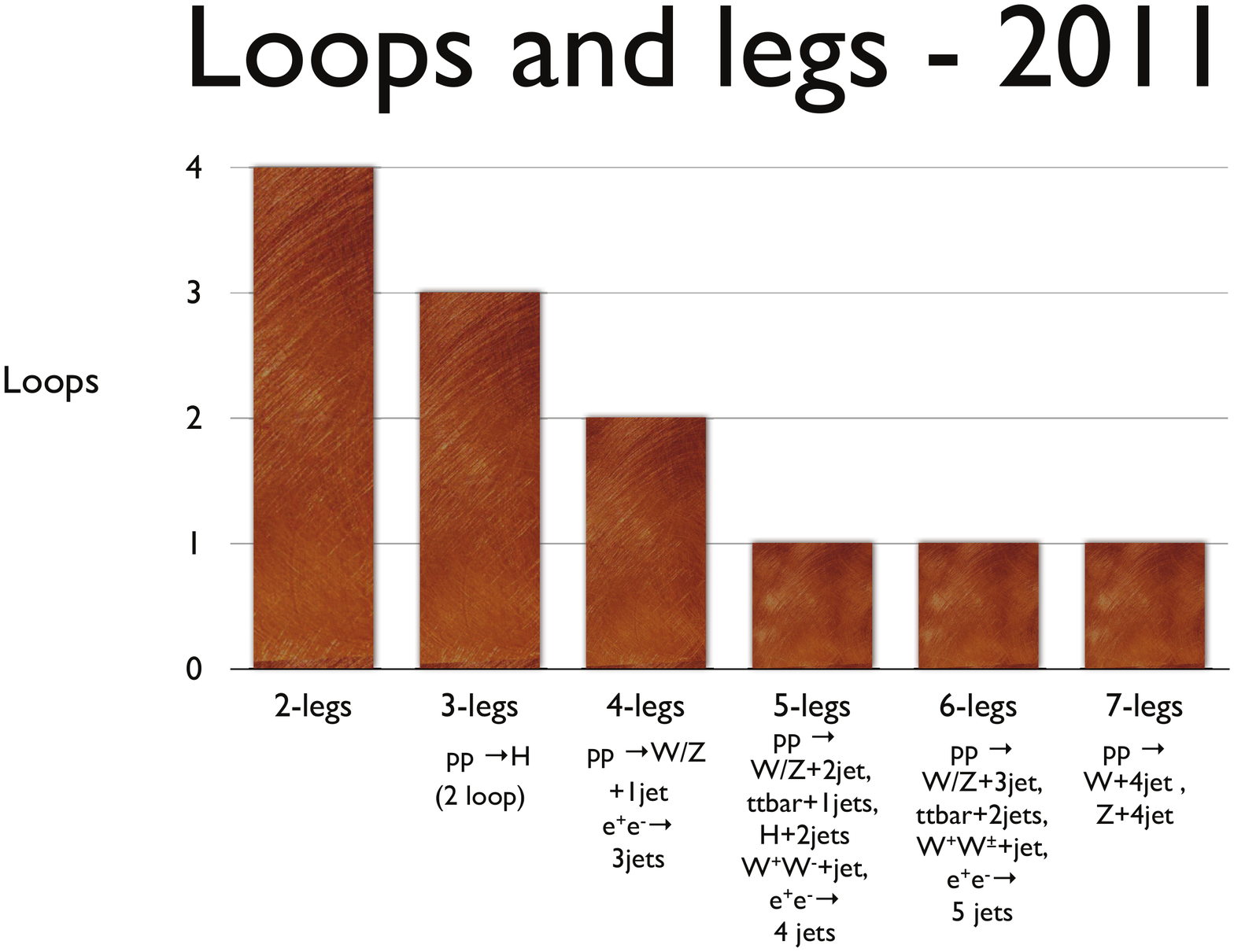}
\end{center}
\caption[]{\small
Loops vs. Legs: the development from 2007 to 2011;
courtesy by R.K. Ellis \cite{RKELLIS}, reprinted with kind permission.}
\end{figure}
%
%
\vspace*{-1mm}
\noindent
Keith Ellis has recently summarized the theoretical progress during the last 4-5 years 
\cite{RKELLIS}, see Fig.~1. This conference adds another entry in column one \cite{CHET1}.
 49 contributions where presented on recent results for multi-leg processes
at NLO, resummations and infrared structure of scattering cross sections, precision calculations 
for low-energy processes, multi-loop corrections, mathematical methods for the calculations in 
quantum field theory, physics at hadron colliders and collider phenomenology, and physics beyond 
the Standard Model. Based on these contributions, I try to give a brief survey on the 
status
of precision calculations currently reached and discuss a few experimental applications.

\section{Multi-Leg Processes} 

\vspace{1mm} 
\noindent
During the last two years quite a series of important $2 \rightarrow 4(5)$ processes have been 
calculated at NLO~:

\begin{tabular}{ll}
$pp     \rightarrow  W^{\pm}(Z,\gamma) + 3~{\rm jets}$,
&  
\cite{Berger:2009ep,Berger:2009zg,KeithEllis:2009bu,Melnikov:2009wh,Berger:2010vm,Campbell:2010cz}
\\
$pp     \rightarrow  W^{\pm}(Z) + 4~{\rm jets}$, 
& \cite{Berger:2010zx,Ita:2011wn}
\\
$pp     \rightarrow  4~{\rm jets}$,
&\cite{Bern:2011ep}
\\
$pp     \rightarrow  t\bar{t} b\bar{b}$, 
& \cite{Bredenstein:2009aj,Bredenstein:2010rs,Bevilacqua:2009zn}
\\
$pp     \rightarrow  t\bar{t} + 2~{\rm jets}$, 
& \cite{Bevilacqua:2010ve,Bevilacqua:2011aa}
\\
$pp     \rightarrow  b\bar{b} b\bar{b}$, 
& \cite{Binoth:2009rv,Greiner:2011mp} 
\\
$pp     \rightarrow  t\bar{t} \rightarrow W^+ W^- b\bar{b}$, 
& \cite{Bevilacqua:2010qb,Denner:2010jp} 
\\
$pp     \rightarrow  W^+ W^+  2~{\rm jets}$, 
& \cite{Melia:2010bm,GOSA1}
\\
$pp     \rightarrow  W^+ W^-  2~{\rm jets}$, 
& \cite{Melia:2011dw}\\
$pp \rightarrow   W \gamma \gamma +~{\rm jet}$, 
& \cite{Campanario:2011ud}
\\
$e^+e^- \rightarrow   \geq 5~{\rm jets}$
& \cite{Frederix:2010ne,Becker:2011vg}~.
\end{tabular}

\noindent
These and a series of related processes are of importance for central measurements and searches
at the LHC, resp. prepare technical steps in the computation of other processes. In these 
calculations a series of computational techniques such as 
MHV amplitudes and their recursion relations, 
cf.~\cite{Parke:1986gb,Berends:1987cv,Berends:1987me,Mangano:1987xk},
cutting techniques, cf.~\cite{CUT1,Ellis:2011cr}, and 
the unitarity method \cite{UNIT} were important, see also \cite{Bern:2008ef}.
\restylefloat{figure}
\begin{figure}[H]
\begin{center}
\includegraphics[scale=0.7]{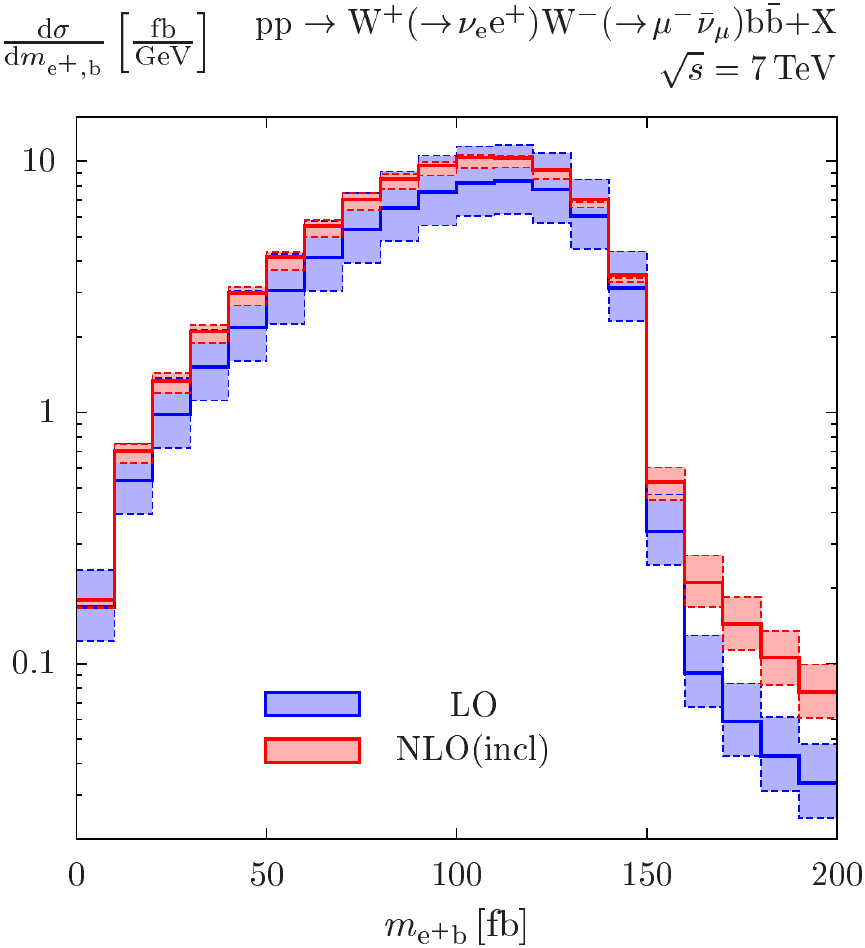}
\end{center}
\caption{\label{fig:wwbb}
\small
Invariant mass $M_{e^+b}$ of the positron-$b$-jet system at
the Tevatron: absolute LO and NLO predictions. The
uncertainty bands describe $m_t/2 < \mu  < 2 m_t$ variations;
from A. Denner, S. Dittmaier, S. Kallweit, and S. Pozzorini, Phys. Rev. Lett. 106 (2011) 052001,
\cite{Denner:2010jp}, \TCop (2011) by the American Physical Society.}
\end{figure}

\noindent
\normalsize
Diagrams contributing to multi-leg processes may contain resonant propagators requiring a 
special treatment \cite{RESO}. Furthermore, the calculation of 5-, 6-, 7-point 
functions may lead to numerical instabilities due to large cancellations, which needs 
special implementations \cite{57point}. 

At this conference reports were given on a variety of multi-leg processes. Hard multi-particle processes at 
NLO QCD were discussed in \cite{VANHAMEREN,BECKER}. The NLO QCD corrections to hadronic $W^+W^-b\bar{b}$ 
production have been calculated in \cite{Denner:2010jp,POZZORINI}. This process is of importance to measure
the $t\bar{t}$ cross section and its background for this channel of final states. 
As an example we show in 
Fig.~\ref{fig:wwbb}\textcolor{black}{~~the invariant
$M_{e^+b}$ mass for the process $pp \rightarrow W^+W^- b \bar{b} +X$.}
Results on multi-boson (+ jet) production were discussed in \cite{ENGLERT,SHIVAJI}. The tensor reduction algorithm 
for one-loop multi-leg Feynman integrals is of importance to obtain stable numerical results. It has been 
derived and implemented up to seven-point functions in \cite{RIEMANN}. New results by the GRACE-collaboration
have been presented in \cite{FUJIMOTO}. The {\tt Golem} and {\tt Samurai} projects have recently been united 
allowing for further advanced calculations \cite{HEINRICH}. The calculation of the process $W^+ W^- jj$ at NLO
was discussed in \cite{MELIA}. A recursive one-loop algorithm for many-particle amplitudes has been presented 
in \cite{MAIERHOEFER}. The signal-background interference in $gg \rightarrow 
H \rightarrow V V$, being not small,
has been discussed in \cite{KAUER}, see also \cite{Campbell:2011cu}. Electro-weak and QCD corrections
to $pp \rightarrow Z^*$ + exponentiation have been implemented in the code {\tt Herwiri2.0} \cite{YOST}.
New aspects in the automation of Standard Model processes in {\tt MadLoop} were described in 
\cite{HIRSCHI}. Alternative NLO subtraction schemes were discussed in \cite{KUBOCZ}. Using the 
Berends-Giele \cite{Berends:1987cv,Berends:1987me} and unitarity method, 1-loop corrections 
to multi jets 
up to 12-14 gluons were 
calculated in \cite{BADGER}. At hadron colliders double parton scattering occurs, i.e. two partonic 
emissions from a single initial state nucleon contribute in a scattering process. For this 
process electro-weak boson production has been calculated in \cite{GAUNT}.

Very efficient multi-leg tools at NLO have been created during the last years, which allow 
the calculation of virtual corrections and real emission. Some of them are equipped with 
hadronic shower algorithms. A number of packages allow to import newly calculated cross sections
for individual reactions in a standardized way. We list a series of codes in alphabetic order.  

\begin{tabular}{ll}
{\tt AutoDipole:}   & Hasegawa, Moch, Uwer, \cite{AUTODIPOLE} \\
{\tt BlackHat:}     & C.~F.~Berger, Bern, Dixon, Febres Cordero, Forde,\\ 
                    & Ita, Kosower  Maitre, \cite{BLACKHAT} \\
{\tt CutTools:}     & Ossola, Papadopoulos, Pittau, \cite{CUTTOOLS}\\
{\tt GOLEM:}        & Binoth, Guillet, Heinrich, Pilon, Reiter, \cite{GOLEM}  \\
{\tt GRACE:}        & Yuasa, Ishikawa, Kurihara, Fujimoto, Shimizu,\\ &
                       Hamaguchi, de Doncker, Kato et al., \cite{GRACE}\\
{\tt Helac/Phegas:} & Czakon, Papadopoulos, Worek, \cite{HELAC} \\
{\tt LoopTools:}    & Hahn et al. + {\tt Feynarts, FormCalc}, \cite{LOOPTOOLS}\\
{\tt MadDipole:}    & Frederix, Greiner, Gehrmann, \cite{MADDIPOLE}\\
{\tt MadFKS:}       & Frederix, Frixions, Maltoni, Stelzer, \cite{MADFKS} \\
{\tt MadLoop:}      & {\tt = CutTools + MadFKS} \\
                       & Hirschi, Frederix, Frixione, Garzelli, Maltoni, Pittau, \cite{MADLOOP}\\
{\tt MCFM:}         & Campbell, R.K. Ellis, Williams, et al., \cite{MCFN}\\
{\tt MC@NLO:}       & Frixione, Webber, \cite{MCNLO}\\
{\tt NGluon:}       & Badger, Biedermann, Uwer, \cite{NGLUON}\\
\end{tabular}
\begin{tabular}{ll}
{\tt NLOJET++:}     & Nagy, Trocsanyi, \cite{NLOJETPP} \\
{\tt POWHEG:}       & Frixione, Nason, Oleari et al., \cite{POWHEG}\\
{\tt Rocket:}       & Giele, Zanderighi, \cite{ROCKET}\\
{\tt Samurai/GoSam:}& Mastrolia, Ossola, Reiter, Tramontano et al. \cite{SAMURAI}\\
{\tt SHERPA:}       & Gleisberg, Krauss et al., \cite{SHERPA} \\
{\tt TeVJet:}       & Seymour, Tevlin, \cite{TEVJET}. 
\end{tabular}

\vspace*{2mm}
\noindent
There are more packages to be released soon, cf.~\cite{Fleischer:2011zz}. We also would like to mention 
recent implementations of NLO parton showers based on unintegrated kernels 
\cite{JADACH}; see also Ref.~\cite{Maestre:2012vp}.
\section{Resummations and Infrared Structure} 

\vspace{1mm} 
\noindent
For many processes the resummation of large logarithms leads to an improvement 
of the theoretical description of differential and inclusive scattering cross sections,
beyond the available fixed order corrections. In some cases these resummations bridge 
between the perturbative and non-perturbative range. A general prescription is required 
to prove the possibility to resum in the particular cases being considered. New insights were provided 
by soft-collinear effective field theory (SCET). A systematic perturbative approach for 
resummations beyond the eikonal-approximation has recently been proposed in \cite{Laenen:2008gt}.

Examples for situations in which large logarithms, $L \gg 1$, need to be resummed are 
\cite{MAGNEA1}~:
\begin{eqnarray}
\text{RGE logs}          &&  \hspace*{1cm} \alpha^k \ln^k(Q^2/\mu^2)     \N\\
\text{High energy logs}  &&  \hspace*{1cm}  \alpha^k \ln^{k-1}(s/t)                   \N\\
\text{Sudakov logs}      &&  \hspace*{1cm} \alpha^k \ln^{2k-1}(1-z),~~z = \mu_1^2/\mu^2_2 \N\\
\text{Coulomb singularity.}  &&                            \N
\end{eqnarray}
Presently resummations are applied for a vast amount of processes: e.g.
the large $x$ behaviour of deep-inelastic structure functions, hadronic final 
states, jet rates, event shapes, the Drell-Yan process, Higgs-boson production, and
heavy quark pair production.
In the case of Sudakov resummation, the general structure of the scattering cross section
in Mellin space reads \cite{Catani:1992ua,MAGNEA1}
\begin{eqnarray}
d\sigma(\alpha_s, N) = H(\alpha_s) \exp\Bigl\{ 
\ln(N) g_1(\alpha_s,N)
+ g_2(\alpha_s,N)
\alpha_s g_3(\alpha_s,N) + \ldots \Bigr\} + O\left(\frac{1}{N}\right)
\end{eqnarray}
and the all-order structure of the perturbative exponent is understood. In other cases,
resummations are considered also for transverse momentum spectra.

\begin{figure}[H]
\begin{center}
\includegraphics[scale=0.9]{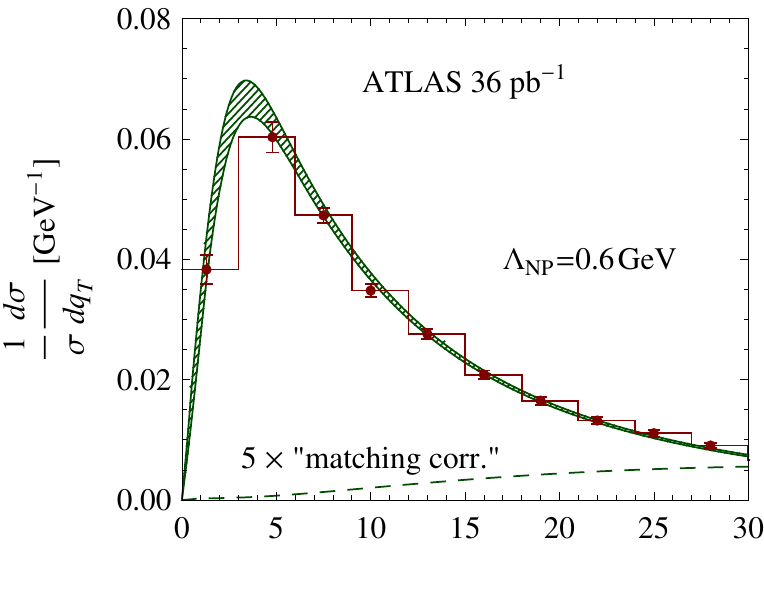}
\end{center}
\caption{\label{fig:DYkt}
\small
The $q_T$--dependence of the resummed Drell-Yan cross section at LHC energies in N$^3$LO
compared to ATLAS-data; from \cite{NEUBERT}, \TCop (2012) Springer Verlag.}
\end{figure}

During this conference a series of contributions were presented on resummations
in high energy processes. In \cite{NEUBERT} the electro-weak gauge-boson production at 
small 
$q_T$ has been studied at N$^3$LL. An illustration is given in 
Fig.~\ref{fig:DYkt} 
\textcolor{black}{comparing} to recent data by ATLAS.
Using the dipole formula, the soft factor of a generic massless high-energy amplitude
has been derived in \cite{DelDuca:2012qg}. This is a promising approach to resum high 
energy logarithms in a systematic way.
The threshold resummation of the total hadronic $t\bar{t}$ cross section at NNLL was 
carried out in \cite{AACHEN,Hoang:2011it}, resumming also the Coulomb singularity  
\cite{AACHEN}.
The resummation of large $x$ terms in semi-inclusive $e^+e^-$-annihilation was 
performed in \cite{LoPresti:2012rg}. 
Amplitude-based resummation in Quantum Field Theories was discussed in 
\cite{Ward:2012sr}.

\restylefloat{figure}
\section{Precision Calculations for Low-Energy Processes} 

\vspace{1mm} 
\noindent
The measurement of fundamental constants and the precise understanding of the associated 
radiative corrections to the fine structure constant $\alpha_{\rm QED}(m_e)$ 
\cite{Jegerlehner:2011mw}, $(g_l-2)$ 
\cite{MELVAI,JEG1,Jegerlehner:2009ry,Aoyama:2012fc,??
ACZ}, Fermi's 
constant $G_F$ and, related to it, precision physics in atomic systems, cf. e.g. 
\cite{Dowling:2009md},
and other pure QED processes, cf. e.g. \cite{Penin:2011aa,CarloniCalame:2011zq}, are of 
great 
importance. These constants form an important part of the present basis of the Standard 
Model. Due to the high experimental accuracy being reached, potential effects due to 
possible extensions of the Standard Model could be revealed comparing with the results of precision 
calculations. At this conference there were reports on the improvement of $a_\mu^{\rm had}$ taking 
into account $\rho$-meson width effects \cite{Jegerlehner:2011ti} and a report on high precision 
luminosity monitors at low energies \cite{Gluza:2012yz}, as well as an improvement of the 
$K_{l3}$ form factor \cite{Gauhar:2012}.

\section{Multi-Loop Corrections} 

\vspace{1mm} 
\noindent
Multi-loop calculations in QCD and QED are progressing reaching the 4- 
resp.\ 5-loop level and include massive 3-loop results. Various steps forward
have been made in case of massless and massive NNLO calculations for $2 \rightarrow 2$
processes in hadronic scattering. All these calculations will allow a much more
precise understanding of important Standard Model processes, and related to it,
the strong coupling constant, parton distributions, the top-quark mass, jet-physics, 
and the origin of particle masses, on the basis of upcoming experimental results at the 
LHC.

At this conference the completed $R_{\rm had}$-ratio to 4-loops 
has been reported \cite{Baikov:2012er} 
\begin{eqnarray}
R(s) &=& 3 \sum_f Q_f^2 \Bigl[ 1 + a_s + a_s^2 (1.986 - 0.1153 N_f) 
         + a_s^3 (6.637 + 1.200 N_f + 0.00518 N_f^2 ) \Bigr] 
\nonumber\\
     & & - \left(\sum_f Q_f^2\right)^2 \left[ 1.2395 a_s^3 
         - a_s^4 \left(17.8277 - 0.57489 N_f \right) \right]~. 
\end{eqnarray}
The $\beta$-function of QED is now known analytically to 5-loop order \cite{CHET1}~\footnote{
See also \cite{CHET2,Kataev:2012rf}.}
\begin{eqnarray}
\beta_{\rm QED} &=& \frac{4}{3} a + 4 a^2 - \frac{62}{9} a^2 - \left(\frac{5570}{243}
+ \frac{832}{9} \zeta_3 \right) a^4
- \left(\frac{195067}{486} + \frac{800}{3} \zeta_3 + \frac{416}{3} \zeta_4 
- \frac{6880}{3} \zeta_5\right) a^5~.
\N\\
\end{eqnarray}
Until very recently the gluon-initiated hadronic inclusive production cross section has only been 
known in the heavy-top approximation. The finite mass effects in the scalar and pseudo-scalar case
have been computed in \cite{HFT,Pak:2011hs}. As has been shown in \cite{Pak:2011hs} the corrections
both in the scalar and pseudo-scalar case are very small for Higgs-masses of $O(120~\GeV)$, but
amount to 9\% (scalar) resp.\ 22\% (pseudoscalar) for masses around $m = 300~\GeV$.
Results of a fully differential NNLO QCD calculations for vector boson and $W$-Higgs production at hadron 
colliders has been presented in \cite{Ferrera:2011bk} including resummations.
The NNLO inclusive Higgs production cross section, including width effects, can be 
calculated with the code {\tt iHixs} \cite{Buehler:2012zf,Anastasiou:2012hx}. 
Comparisons for a wider class of parton distribution functions are provided. In higher 
order calculations one central problem consists in disentangling of overlapping 
singularities. One way consists in special non-linear Feynman parameter mapping 
\cite{Herzog:2012ss,Anastasiou:2010pw}, which allow the numeric calculations of the
coefficients in the $\ep$-expansion. The method goes back to Hamberg and van Neerven 
\cite{HAMBERG}. The degree of non-linearity of the corresponding representation may 
limit an eventual analytic calculation, which, however, is not always intended. 
Non-planar massive double boxes have been calculated based on the code {\tt Reduze2} 
\cite{vonManteuffel:2012yz}. A related numerical calculation of two-loop box diagrams 
has been carried out in \cite{FUJIMOTO}. Massive Wilson coefficients in the asymptotic 
region $Q^2 \gg m^2$ factorize into massive operator matrix elements and massless 
Wilson coefficients \cite{Buza:1995ie}. While this has been shown by an explicit 
calculations to work at $O(\alpha^2)$ in case of massless external lines, a calculation 
in case of massive on-shell external lines has only been accomplished recently 
\cite{Blumlein:2011mi} calculating the $O(\alpha^2)$ massive OMEs 
contributing to the process $e^+e^- \rightarrow \gamma^*/Z^*$. Here, the logarithmic 
terms at $O(\alpha^2)$ yield the desired result, which is not the case for the constant 
terms, unlike at $O(\alpha)$, needing further investigation. Massive OMEs with 
massless external lines have been calculated to 3-loops in QCD 
\cite{Ablinger:2012qj,Ablinger:2011pb,Ablinger:2010ty,HQ12} at general values of the Mellin 
variable $N$ generalizing results obtained for fixed moments in \cite{Bierenbaum:2009mv}, 
resp.\ fixed moments in case of two different fermion masses. Currently 
techniques are available to calculate the bubble- and ladder topologies. The summation 
methods and codes being used in the calculation have been improved essentially 
\cite{Blumlein:2012hg}.

The knowledge of jet-production cross sections at LHC energies at NNLO is of 
central importance. This accuracy is required for sensible QCD tests, further
constraints on the gluon and sea-quark densities, and a correct background description 
for various processes. Furthermore, a much deeper insight into QCD-scattering processes 
is obtained and various valuable new computation technologies are developed for these
calculations. The same techniques allow the calculation of a series of other
processes like $pp \rightarrow \gamma + {\rm jet}, 2 \gamma,  V + {\rm jet}, 
VV, H^0 + {\rm jet}$. In earlier works the NNLO corrections to $e^+e^- \rightarrow~{\rm 
3 jets} + X$ have been computed, cf.~\cite{ZURICH1,Weinzierl:2009nz}. Currently the 
calculation of the NNLO corrections to $pp \rightarrow {\rm 2~jets}$ are underway, with 
contributions from different groups \cite{Currie:2011nb,NNLOjet}.
Also the knowledge of the NNLO corrections to $ep \rightarrow {\rm 2~jets} + X$ is 
of great importance. The available high precision data measured at HERA 
\cite{HERAJET,Grindhammer:2011ur,Kogler:2011hw} will allow another precision 
measurement of the $\alpha_s(M_Z^2)$, still suffering from a large theory-error at NLO.

Another important process at the LHC is $pp \rightarrow t\bar{t}$. Presently the 
NNLO corrections are known in approximate form based on NLO + threshold resummation and are 
available in different codes \cite{Moch:2008qy,Aliev:2010zk,Czakon:2011xx}. Recently 
also the Coulomb corrections have been include \cite{AACHEN,Beneke:2012eb}. The
calculation of the complete corrections, forming a challenging task, are presently in 
progress with contributions from different groups, see \cite{TTBAR}.

\section{Mathematical Methods in Feynman Diagram Calculations}

\vspace{1mm}
\noindent
Depending on the number of loops, legs, and scales involved in the corresponding 
problem the calculations can be either performed analytically, semi-analytically, or
numerically. Numerical problems are solved using the languages {\tt Fortran, C} and 
{\tt C++} at large farms. Many analytic calculations are based on {\tt FORM} in its 
versions {\tt tform} and {\tt parform} \cite{FORM}, applying multiple threads on
main frames or parallelizing to different processors in farms. Other computer 
algebra codes are written in {\tt maple} \cite{MAPLE}, {\tt mathematica} \cite{MATHEMATICA}, 
or {\tt ginac} \cite{ginac}. Currently typical main frames are equipped with 200-300 Gbyte
RAM and fast discs, which are about 10-20 times larger. There are problems for which an 
amount of 2 peta terms need to be processed \cite{MZV}, which requires run times of the 
order of one CPU year, to quote an example.

There are general tools for the generation of Feynman diagrams like {\tt QGRAF} 
\cite{QGRAF} and systematic ways to the Feynman parameter integrals, like graph
polynomials \cite{GRAPHPOL}.

Comparing various calculations the general observation is made that the 
results have common basis
representations, re-appearing in the solution of many different problems \cite{MVV1,JR1,BKKS}. In 
case of 0-scale problems, such as moments for anomalous dimensions,
or expansion coefficients of the $\beta$-function, these are special numbers, like
multiple zeta values \cite{MZV}, or corresponding values of iterated integrals at 
argument $x=1$ \cite{ITINT} over some special alphabets, up to those being generated 
by cyclotomic polynomials and elliptic integrals, \cite{HPOL,hpl,CYCL,ELL}. Single 
scale 
quantities can be expressed in terms of harmonic sums \cite{summer,HSUM1}, harmonic 
polylogarithms \cite{HPOL}, hyperlogarithms \cite{HYPL}, cyclotomic polylogarithms 
\cite{CYCL}, generalized harmonic sums \cite{GENS1,GENS2} and other extensions. General 
structures in case of two- and more scale problems at 2 loops and higher 
have not been studied systematically yet, but do certainly exist. 
 
Integration and summation methods to solve 0-scale problems in the massless and massive
case have been standardized in several packages like {\tt MINCER} \cite{MINCER},
Baikov's method \cite{BAIKOV}, {\tt MATAD} \cite{MATAD}, {\tt qexp} \cite{QEXP},
a code for 4-loop vacuum-bubble master integrals \cite{Schroder:2005va}, and
{\tt Sigma} \cite{SIGMA}. One may use PSLQ-based \cite{PSLQ} methods \cite{DB,MZV,LSS}
to guess the corresponding quantity based on highly precise numerical values, and apply
hyperlogarithms\footnote{Hyperlogarithms are distinct from the usual iterated integrals 
with multi-linear denominator functions, since the remainder variables are not constant 
but may be integrated over.} at infinite argument \cite{HYPL}. 

In the calculation of 1-dimensional and higher dimensional quantities various methods 
play an important role~: 
\begin{itemize}
\item
{\it Integration by parts.} Gau\ss{}' theorem \cite{IBP} and relations implied by 
Lorentz-invariance
allow to reduce Feynman integrals to so-called master integrals. There are various 
implementations
of the corresponding algorithms \cite{Laporta:2001dd}, including the recent public codes {\tt Air} 
\cite{ANAST}, {\tt FIRE} \cite{FIRE} and  {\tt REDUZE} 
\cite{REDUZE1,vonManteuffel:2012yz}.
\item
{\it Sector decomposition.} The decomposition of the integration range of individual 
Feynman parameter integrals allows the extraction of their singularities \cite{HEPP,BH,NS,ABD}. 
Corresponding codes are {\tt FIESTA} \cite{SST},
{\tt sector\_decomposition} \cite{BW1},  {\tt CSectors} 
\cite{GKRY}, {\tt SecDec} \cite{CH}. 
\item
{\it Mellin-Barnes integrals.} Mellin-Barnes transformations of Feynman parameter 
integrals \cite{MELB,BL,UD} allow the $\varepsilon$-expansion and the numeric, or in 
certain cases also the analytic, calculation of the expansion coefficients. Corresponding 
codes are {\tt MB.m, MBasymptotics.m} \cite{MB}, {\tt barnesroutines.m} \cite{KOSOWER}, 
{\tt AMBRE.m} \cite{AMBRE}, {\tt MBresolve} \cite{SS}.
\item
{\it Use of differential equations.} Master integrals can be often calculated 
using differential equations \cite{CCLR,GR,CCR}. There are numerous applications of 
this method.
\item
{\it Generalized hypergeometric and related functions.} Feynman parameter 
integrals at 2-loops and in some cases at 3-loops can be represented by these higher
transcendental functions \cite{KALM,ABHKSW,Ablinger:2012qj,HQ12,BKK}. Various packages 
for the 
expansion in
the dimensional parameter $\varepsilon$ around integer and half-integer values exist, 
as {\tt HypExp, HypExp2} and {\tt HyperDire} \cite{HM,KALM,WZ}. 
\item
{\it Difference equations.}  Feynman diagrams in Mellin space are related by difference equations 
\cite{Blumlein:2009ta,Blumlein:2010zv}. This method was applied systematically e.g. in
in \cite{Moch:1999eb,Moch:2004pa,Vogt:2004mw,MVV1,Bierenbaum:2008yu,BKKS}. 
\item
{\it Summation methods.} Feynman parameter integrals can be transformed into multiply nested sums
over hypergeometric terms. Sums of this kind can be solved in difference- and product 
fields. 
Powerful algorithms have been implemented in the package {\tt Sigma} by C. Schneider \cite{SIGMA}.  
Related to this, the application of multi-sum algorithms may be useful 
\cite{WEGSCHAIDER,Blumlein:2010zv}. 
\item
{\it Recurrences from moments.} If a sufficiently large number of moments for a recursive quantity
can be generated, one may find its recurrence using the so-called guessing method \cite{MKguess}.
It works reliably for very large systems and e.g. allows to reconstruct the 3-loop anomalous dimensions
and massless Wilson coefficients. Here up to 5114 moments would be required for individual color factors
and the recurrences obtained are of order 35 and degree $\sim 1000$. They can be solved using available
summation technologies \cite{BKKS}.
\item
{\it Integration and holonomic functions.}  
In case of holonomic functions the associated multi-variate difference or differential equations
can in principle be obtained using the Almkvist-Zeilberger algorithm \cite{ALMZEI}. Implementations are
given in \cite{CK,HarmonicSums}.
\end{itemize}

\noindent
Relations between harmonic sums \cite{summer,HSUM1,Blumlein:2003gb}
and their generalizations \cite{GENS1,GENS2}, 
resp. harmonic polylogarithms (and generalizations), including cyclotomic harmonic sums, polylogarithms 
and their generalizations \cite{CYCL} are encoded in  packages like
{\tt summer} \cite{summer},
{\tt harmpol} \cite{HPOL},
{\tt hpl} \cite{hpl},
{\tt nestedsums} \cite{nestedsums,VW},
{\tt HPL} \cite{HAPL},
{\tt Xsummer} \cite{Moch:2005uc},
{\tt HarmonicSums} \cite{HarmonicSums}, and
{\tt CHAPLIN} \cite{CHAPLIN}. A large data base for Euler-Zagier values was given in the 
multiple zeta value data mine \cite{MZV}. Analytic continuations of harmonic sums to complex
arguments $N$ are given in \cite{ANCONT,Blumlein:2009ta,CYCL,GENS2}.
 
\newpage
\section{Parton Distributions for the LHC}
\subsection{NNLO PDFs}

\begin{figure}[H]
\begin{center}
\includegraphics[scale=0.5,angle=0]{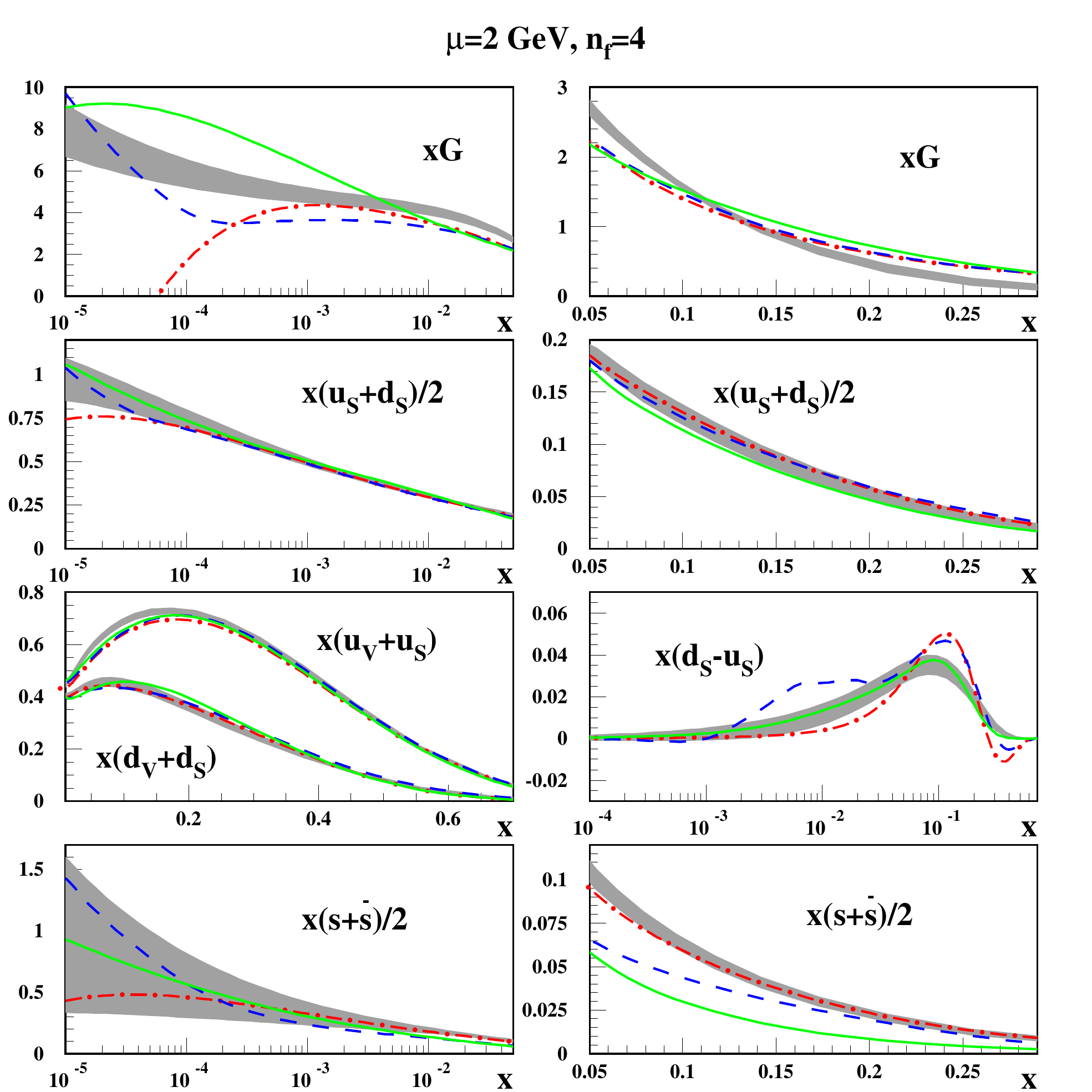}
\end{center}
\caption{
\label{fig:pdf}
\small
The 1 $\sigma$ band for the 4-flavor NNLO ABM11 PDFs \cite{Alekhin:2012ig} at the scale 
of $\mu = 2$~GeV versus $x$ (shaded area) compared with the PDFs obtained by other 
groups. Solid lines: JR09 \cite{JimenezDelgado:2008hf}, dashed dots: MSTW08 
\cite{Martin:2009iq}, 
dashes: NN21 \cite{Ball:2011us}; from~\cite{Alekhin:2012ig}.
}
\end{figure}

\vspace{1mm}
\noindent
Let us now discuss some precision measurements in QCD at high energy colliders. The physics
at the hadron colliders Tevatron and LHC depends very sensibly on the detailed knowledge of 
the parton distribution functions (PDFs). During the last decade they were improved steadily.
At present they are determined with 3-loop accuracy from the world deep-inelastic and
other precise hard scattering data. The five groups AB(K)M \cite{Alekhin:2009ni,Alekhin:2012ig}, 
HERAPDF \cite{HERAPDF}, JR \cite{JimenezDelgado:2008hf}, MSTW08 \cite{Martin:2009iq}
and NNPDF \cite{Ball:2011us} have carried out NNLO analyses and CTEQ will release NNLO results 
soon. Precision determinations of PDFs have to refer to consistent precision data. Here the use 
of the combined H1 and ZEUS data \cite{HERA:2009wt} is rather essential. The usual statistical 
measure $(\Delta \chi^2 = 1)$, cf. \cite{Alekhin:2009ni,Alekhin:2012ig,Ball:2011us}, should 
be 
used, treating the systematic errors separately. Subsamples of precision data should reflect 
appropriately with their parameters within the global fit. 
In Fig.~\ref{fig:pdf} \textcolor{black}{the results
of present PDF-fits are compared.} 
\restylefloat{figure}

\noindent
While the valence and sum of the light sea quark distributions agree rather well, there are still
big differences in the gluon distribution, the difference of the light sea quark
distributions and in case of the strange sea. These differences have an impact on the predictions
of different scattering cross sections at the LHC. In Ref.~\cite{Alekhin:2012ig}  a series of 
reasons for these differences were analyzed and the discussions between the fitting groups  
on further improvements are ongoing.
\subsection{$\alpha_s(M_Z^2)$}

\vspace{1mm}
\noindent
The determination of the strong coupling constant $\alpha_s(M_Z^2)$ using various precision 
measurements and perturbative precision calculations resp. lattice simulations has been discussed
recently in detail in Ref. \cite{Bethke:2011tr}. Due to the high accuracy we will compare only
the determinations at NNLO and higher, see~Table~\textcolor{black}{\ref{tab:alph}.} 

\begin{table}[h!]\centering
\renewcommand{\arraystretch}{1}
\begin{tabular}{|l|l|l|}
\hline
\multicolumn{1}{|c|}{ } &
\multicolumn{1}{c|}{$\alpha_s({M_Z})$} &
\multicolumn{1}{c|}{  } \\
\hline
BBG      & $0.1134~^{+~0.0019}_{-~0.0021}$
         & {\rm valence~analysis, NNLO}  \cite{Blumlein:2004ip,Blumlein:2006be}
\\[0.5ex]
BB       & $0.1132  \pm 0.0022$
         & {\rm valence~analysis, NNLO}  \cite{BBprep}
\\
GRS      & $0.112 $ & {\rm valence~analysis, NNLO}  \cite{Gluck:2006yz}
\\
ABKM           & $0.1135 \pm 0.0014$ & {\rm HQ:~FFNS~$n_f=3$} \cite{Alekhin:2009ni}
\\
ABKM           & $0.1129 \pm 0.0014$ & {\rm HQ:~BSMN-approach}
\cite{Alekhin:2009ni}
\\
JR       & $0.1124 \pm 0.0020$ & {\rm
dynamical~approach} \cite{JimenezDelgado:2008hf}
\\
JR       & $0.1158 \pm 0.0035$ & {\rm
standard~fit}  \cite{JimenezDelgado:2008hf}
\\
ABM11            & $0.1134\pm 0.0011$ &   \cite{Alekhin:2012ig}\\
MSTW & $0.1171\pm 0.0014$ &  \cite{Martin:2009bu}     \\
NN21 & $0.1173\pm 0.0007$ &  \cite{Ball:2011us}     \\
CT10 & $0.118\phantom{0} \pm 0.005$  &  \cite{CTEQ12} \\
\hline
Gehrmann et al.& {{$0.1153 \pm 0.0017 \pm 0.0023$}} & {\rm
$e^+e^-$~thrust}~\cite{Gehrmann:2009eh}
\\
Abbate et al.& {{$0.1135 \pm 0.0011 \pm 0.0006$}} & {\rm
$e^+e^-$~thrust}~\cite{Abbate:2010xh}
\\
\hline
3 jet rate   & $0.1175 \pm 0.0025$ & Dissertori et al. 2009 \cite{Dissertori:2009qa}\\
Z-decay      & $0.1189 \pm 0.0026$ & BCK 2008/12  (N$^3$LO) 
\cite{Baikov:2008jh,Baikov:2012er}\\
$\tau$ decay & $0.1212 \pm 0.0019$ & BCK 2008               \cite{Baikov:2008jh}\\
$\tau$ decay & $0.1204 \pm 0.0016$ & Pich 2011              \cite{Bethke:2011tr}\\
$\tau$ decay & $0.1180 \pm 0.0008$ & Beneke, Jamin 2008     \cite{Beneke:2008ad}
\\
\hline
lattice      & $0.1205 \pm 0.0010$ & PACS-CS 2009 (2+1 fl.) \cite{Aoki:2009tf} \\
lattice      & $0.1184 \pm 0.0006$ & HPQCD 2010             \cite{McNeile:2010ji} \\
lattice      & $0.1200 \pm 0.0014$ & ETM 2012 (2+1+1 fl.)   \cite{Blossier:2012ef}
\\
\hline
BBG & $0.1141~^{+~0.0020}_{-~0.0022}$
& {\rm valence~analysis, N$^3$LO$(^*)$}  \cite{Blumlein:2006be}            
\\[0.5ex]
BB & $0.1137 \pm 0.0022$
& {\rm valence~analysis, N$^3$LO$(^*)$}  \cite{BBprep}            \\
\hline
{world average} & {$
0.1184 \pm 0.0007$  } & \cite{Bethke:2009jm} (2009)
\\
                & {$
0.1183 \pm 0.0010$  } & \cite{Bethke:2011tr} (2011)
\\
\hline
\end{tabular}
\renewcommand{\arraystretch}{1}
\caption{
\small
\label{tab:alph}
Summary of recent NNLO QCD analyses of the DIS world data, supplemented by related measurements
using other processes; from \cite{Alekhin:2012ig}.}
\end{table}

\noindent
Flavor non-singlet analyses of 
the DIS world data were performed in~\cite{Blumlein:2006be,Gluck:2006yz,BBprep}, with an accuracy 
of $\Delta \alpha_s(M_Z) \simeq 2\%$. The difference between the value at N$^3$LO$^*$ and NNLO 
amounts to $\sim 0.0007$ indicating the size of remaining uncertainty. A difference of $\Delta 
\alpha_s(M_Z) = 0.0006$ due to the treatment of the heavy-flavor corrections was observed.
These uncertainties signal the typical theory errors remaining at the present level of description.
The combined flavor non-singlet and singlet analyses 
\cite{JimenezDelgado:2008hf,Alekhin:2009ni,Alekhin:2012ig} 
obtained quite similar values. 
The inclusion of Tevatron jet data, cf.~\cite{Alekhin:2011cf},
although only a NNLO$^*$ analysis, alters this values at most to $\alpha_s(M_Z) = 0.1149 \pm 
0.0012$. Re-analyzes have to be performed as soon as the NNLO corrections become available.
Low values of $\alpha_s(M_Z)$ have also been found in the analysis of thrust in 
$e^+ e^-$-annihilation in~\cite{Abbate:2010xh,Gehrmann:2009eh}.
Larger central values of $\alpha_s(M_Z)$ at NNLO are reported by MSTW \cite{Martin:2009bu} and 
NN21 \cite{Ball:2011us}. These fits include a much broader set of hadronic scattering data in the
analysis. A detailed discussion of sources causing these higher values has been  given in 
Ref.~\cite{Alekhin:2012ig}. A (preliminary) central value of $\alpha_s(M_Z)$ reported by
CT10~\cite{CTEQ12} is similar to MSTW and NN21 at NNLO,
although accompanied by a rather large uncertainty of $\Delta \alpha_s(M_Z) = 0.0050$.
Larger central values for $\alpha_s(M_Z)$ are obtained for the 3-jet rate in $e^+e^-$ 
annihilation~\cite{Dissertori:2009qa} at NNLO
and for the $Z$-decay width at N$^3$LO \cite{Baikov:2008jh}.
The present $\alpha_s(M_Z)$ values at NNLO extracted from $\tau$-decays  
vary between 0.1212 and 0.1180 \cite{Baikov:2008jh,Bethke:2011tr,Beneke:2008ad}.
$\alpha_s(M_Z^2)$ was also determined in different lattice simulations.

Despite the high precision on $\alpha_s(M_Z^2)$ reached for different observables there is 
no 
consensus yet reached on the value of $\alpha_s(M_Z^2)$. The systematics between different 
measurements has still to be further understood and further precision data are needed.
Ideal measurements could be performed at the Giga--$Z$ option at a future linear collider.

\subsection{$W^{\pm}$ and $Z$-boson production}

\vspace{1mm}
\noindent
\noindent
The inclusive $W^{\pm}$ and $Z$-boson production cross sections belong to the standard candles
at the LHC, for which NNLO predictions have been calculated, cf.~\cite{Alekhin:2010dd}. The cross 
sections and their 
\restylefloat{figure}
\begin{figure}[H]
\begin{center}
\includegraphics[scale=0.35,angle=0]{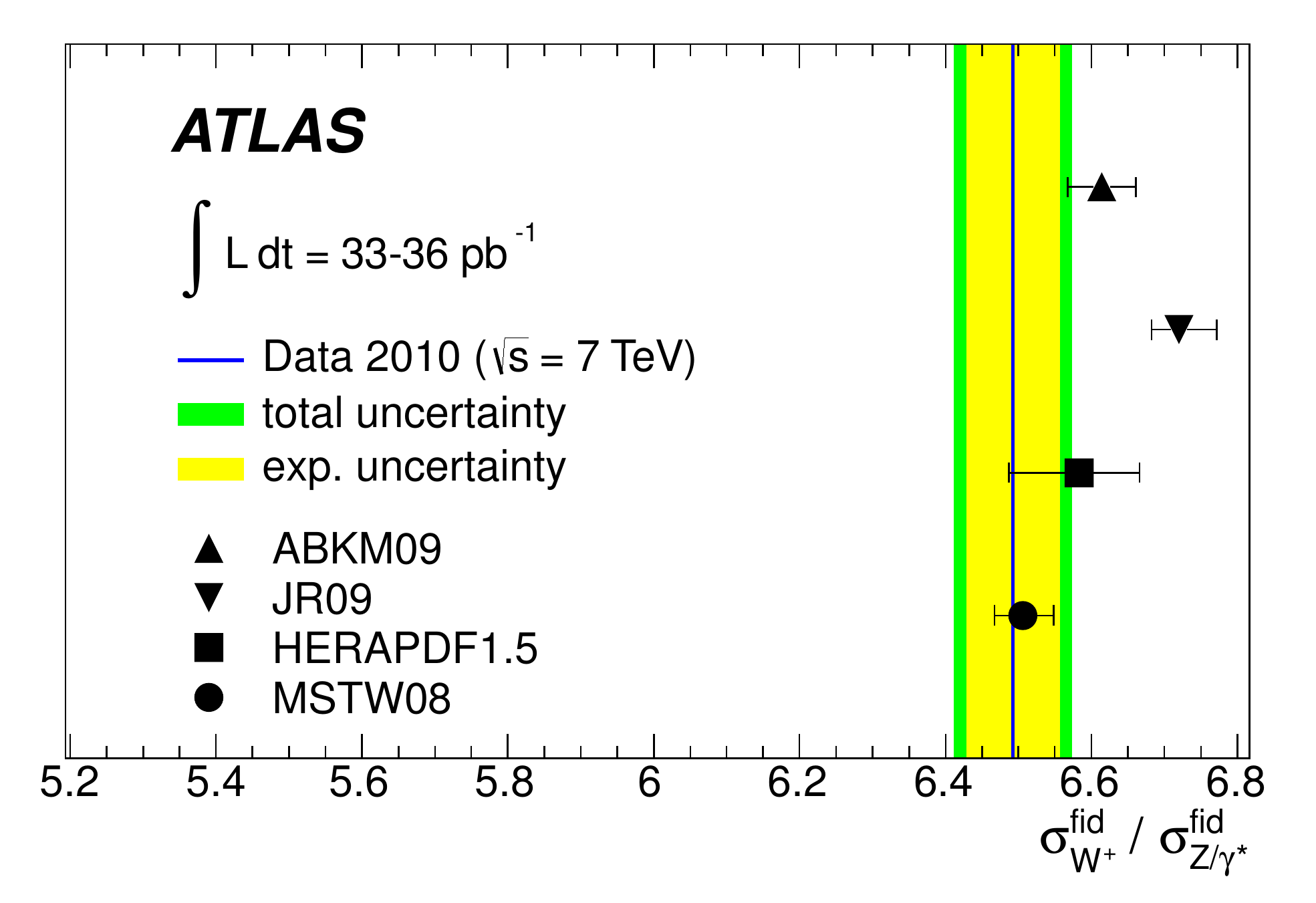}
\includegraphics[scale=0.35,angle=0]{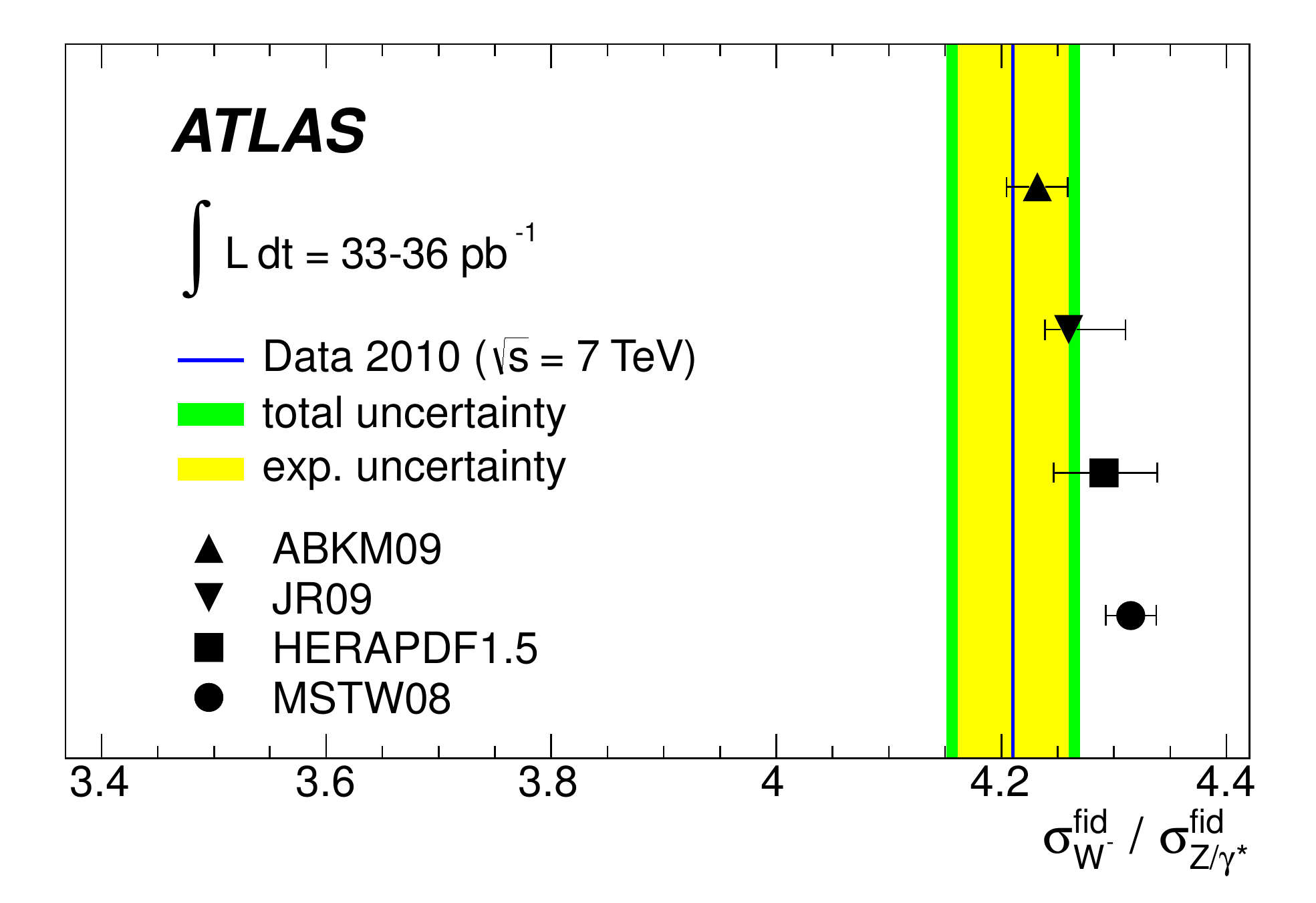}
\end{center}
\caption{
\label{fig:WZrat1}
\small
Measured and predicted fiducial cross section ratios, $\sigma(W^+)/\sigma(Z^0)$ 
(left) and $\sigma(W^-)/\sigma(Z^0)$ (right). The experimental uncertainty (inner yellow band) 
includes the experimental systematic errors. The total uncertainty (outer green band) includes
the statistical uncertainty and the small contribution from the acceptance correction. The 
uncertainties of the ABKM \cite{Alekhin:2009ni}, JR \cite{JimenezDelgado:2008hf} and MSTW08 
\cite{Martin:2009iq} predictions are given by the PDF uncertainties considered to correspond to 
68 \% CL and their correlations are derived from the eigenvector sets. The results for HERAPDF 
comprise all three sources of uncertainty of that set; from~\cite{Aad:2011dm}, 
Phys.\ Rev.\ D {\bf 85} (2012) 072004, \TCop (2012) by the American Physical Society.}
\end{figure}

\noindent
ratios $\sigma(W^\pm)/\sigma(Z^0)$ were measured with a high accuracy at 
ATLAS \cite{Aad:2011dm}, CMS \cite{CMS:2011aa} and LHCb, \cite{Aaij:2012vn}. The cross 
section ratios are already nearly as precise as the theoretical predictions only taking into 
account the PDF-errors, see Fig.~\ref{fig:WZrat1}\textcolor{black}{, since the luminosity 
errors and part of 
other systematic errors cancel. The different predictions do well agree with the measurements and 
a fit of these data will improve the present accuracy in the sea quark sector.}

\begin{figure}[H]
\begin{center}
\includegraphics[scale=0.35,angle=0]{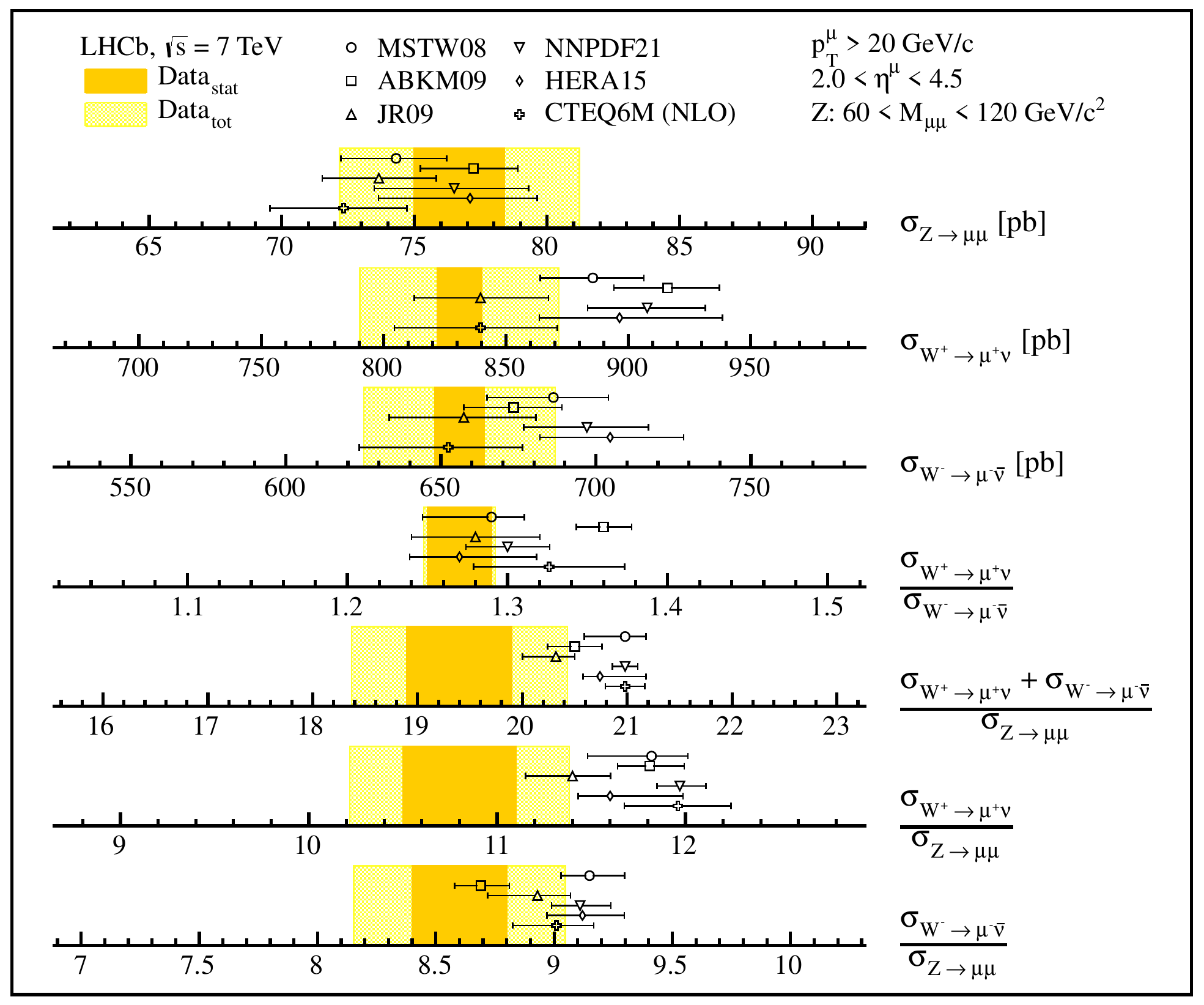}
\end{center}
\caption[]{
\label{fig:WZrat2}
\small
Measurements of the $Z, W^+$ and $W^-$ cross-section and ratios, data are shown as bands 
with 
the statistical (dark shaded/orange) and total (light hatched/yellow) errors.
The measurements are compared to NNLO and NLO predictions with different PDF sets
for the proton (ABKM \cite{Alekhin:2009ni}, JR \cite{JimenezDelgado:2008hf}  and  MSTW08 
\cite{Martin:2009iq}, NNPDF \cite{Ball:2010de}, and CTEQ \cite{Nadolsky:2008zw}),
shown as points with error bars. The PDF uncertainty, evaluated at the
68\% confidence level, and the theoretical uncertainties are added in quadrature to obtain
the uncertainties of the predictions; from  \cite{Aaij:2012vn} with kind permission of the 
LHCb collaboration.}
\end{figure}

The LHCb measurements \cite{Aaij:2012vn} shown in 
Fig.~\ref{fig:WZrat2}\textcolor{black}{, due to its forward} 
kinematics, are sensitive to the quark and anti-quark distributions at smaller values of $x$.
Again, these data will improve the sea-quark distributions. 

\restylefloat{figure}

\subsection{Jet production}

\begin{figure}[H]
\begin{center}
\includegraphics[scale=0.70,angle=0]{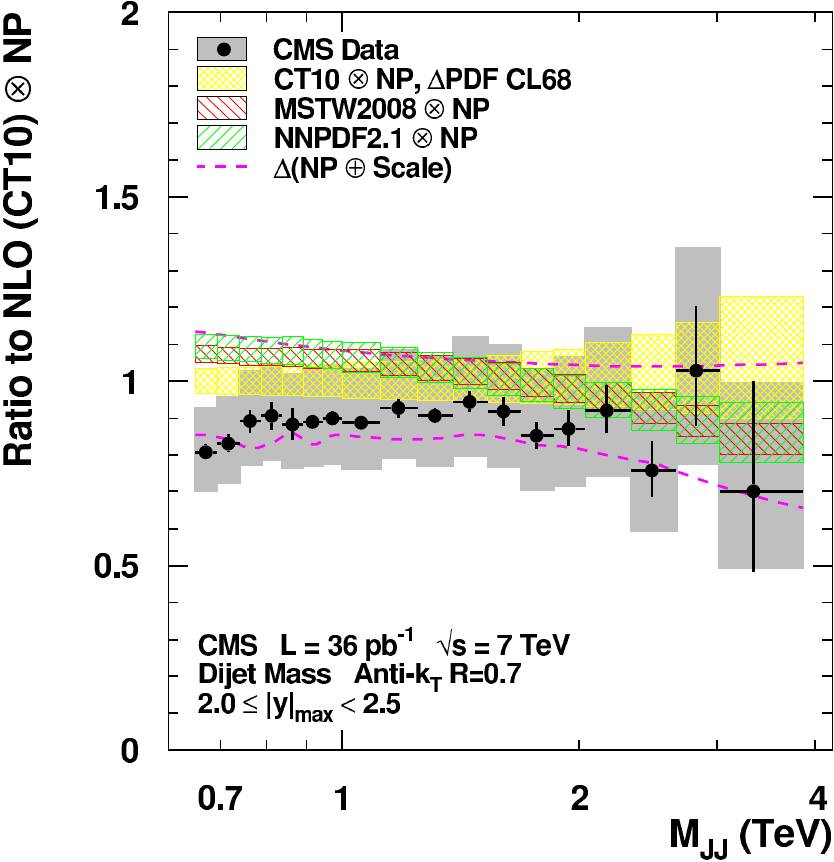}
\includegraphics[scale=0.70,angle=0]{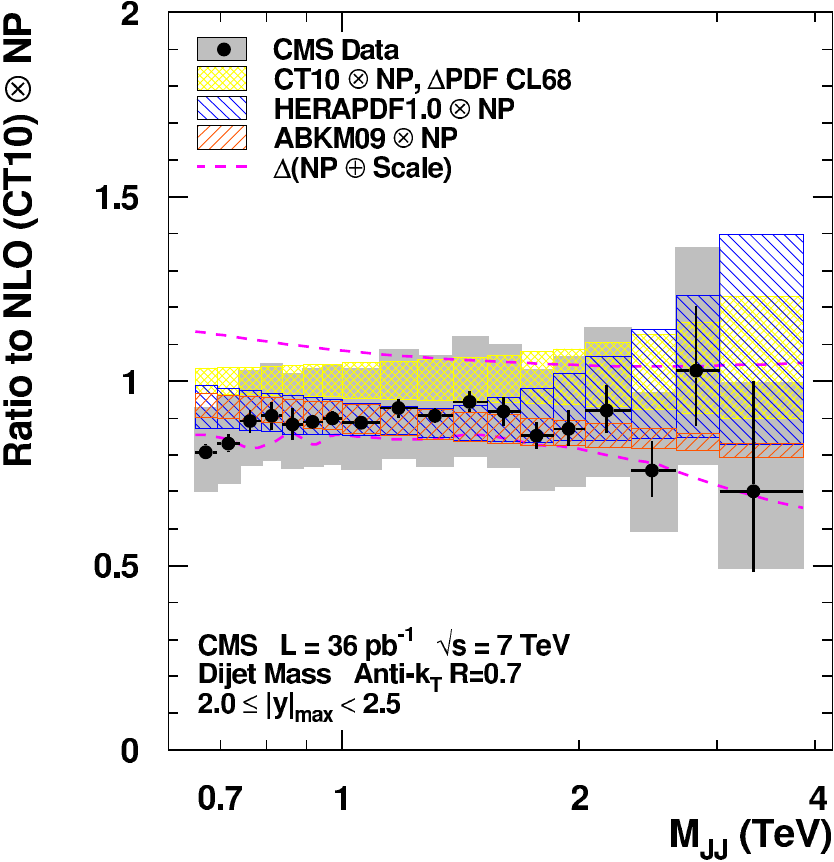}
\end{center}
\caption[]{
\small
\label{fig:JET}
CMS dijet mass data for $2.0 < |y|_{max} < 2.5$ are presented vs. $M_{jj}$ with statistical 
(error bars) as well as systematic uncertainties (grey band) as ratio to NLO using the 
CT10 PDFs \cite{CT10}. Additional prediction are shown using the MSTW2008 and NNPDF2.1 (left) 
and the HERAPDF1.0 \cite{HERAPDF} and ABKM09 PDFs \cite{Alekhin:2009ni}  (right). PDF 
uncertainties are 
displayed as colored bands. Common theoretical uncertainties from scale choices and 
non-perturbative (NP) 
corrections are indicated by dashed magenta lines; from \cite{RABBERTZ} with kind permission.
}
\end{figure}

\vspace{1mm}
\noindent
The jet cross sections at LHC are particularly sensitive to the gluon distribution and the
value of $\alpha_s(M_Z^2)$. Di- and multi-jet final states allow for a better definition of hard 
scales involved and are expected to allow for more direct comparisons to perturbative
predictions. CMS has compared their jet distributions to predictions based on different sets of 
NLO parton densities \cite{RABBERTZ} in Fig.~\ref{fig:JET}\textcolor{black}{.}
The dijet mass data show a sensitivity w.r.t. to the different parton distributions at the level 
of 20--30\% at NLO. The systematic and scale variation errors are still large. The comparison
shows that MSTW \cite{Martin:2009iq} and NNPDF \cite{Ball:2011mu} give predictions a bit higher 
than the 
data, while 
HERAPDF \cite{HERAPDF} and ABKM09 \cite{Alekhin:2009ni} are closer to the central values. The 
data are 
gluon-dominated and one may expect good constraints from these and other jet data on the gluon 
distribution in the future.

\subsection{Higgs Search}

\begin{figure}[H]
\begin{center}
\includegraphics[scale=0.4,angle=0]{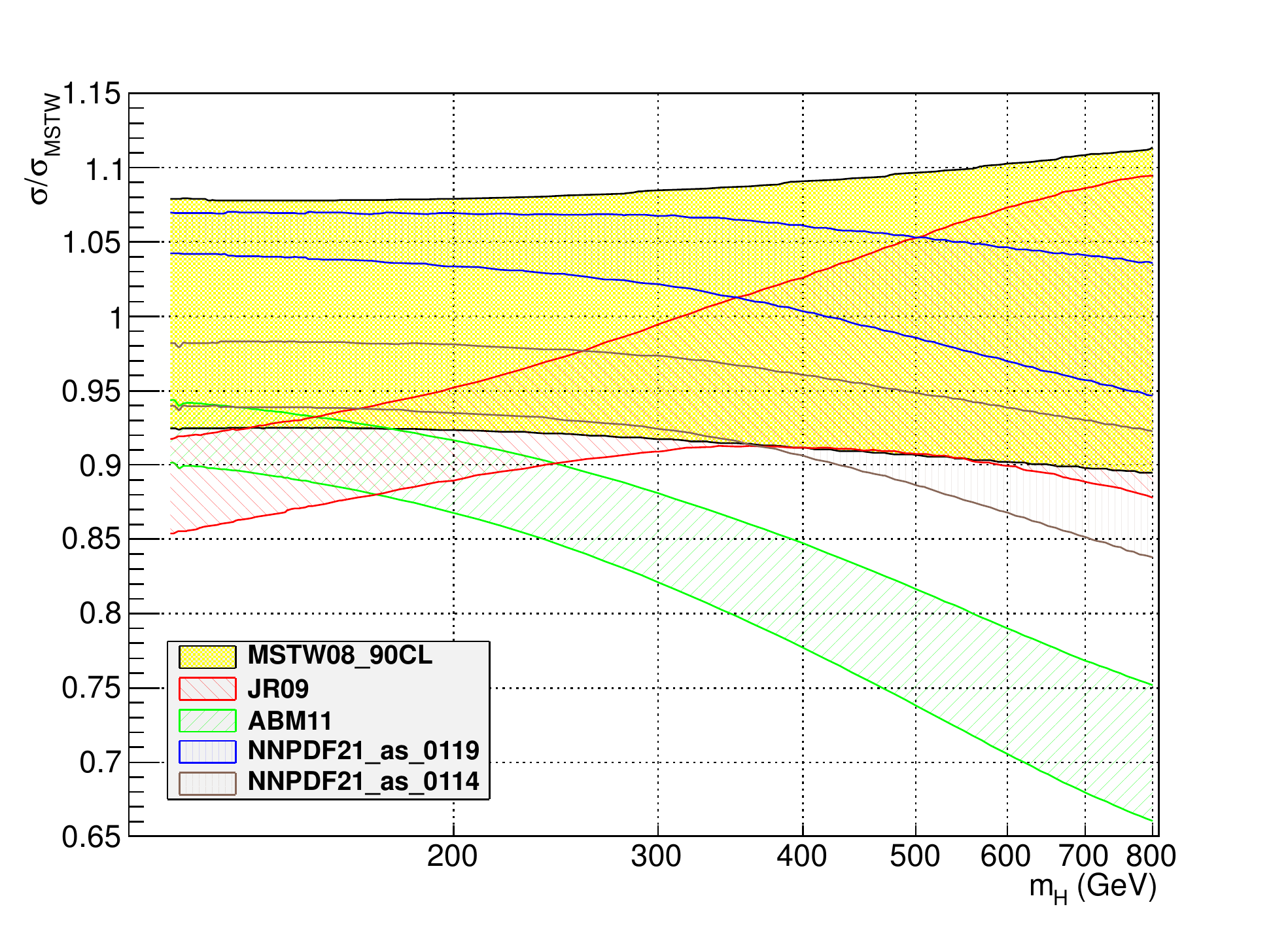}
\end{center}
\caption[]{\label{fig:HIGxs}
\small
The uncertainty of the Higgs cross-section due to the different sets of parton 
distribution functions ABM11 \cite{Alekhin:2012ig}, JR09 \cite{JimenezDelgado:2008hf}, MSTW08 
\cite{Martin:2009iq}, and NN21 \cite{Ball:2011us}; from \cite{Anastasiou:2012hx} 
\TCop (2012) Springer Verlag.}
\end{figure}

\vspace{1mm}
\noindent
The search for the Higgs boson(s) of the Standard Model and of its possible extensions 
is one of the central tasks of the LHC experiments. The main production process, $gg \rightarrow 
H^0$, depends on the PDFs and $\alpha_s(M_Z^2)$ like $ \alpha_s^2 xG(x) \otimes xG(x)$, where
$xG(x)$ denotes the gluon momentum distribution and $\otimes$ the Mellin convolution.
The remaining uncertainties both in the PDFs and in $\alpha_s(M_Z^2)$ propagate into the current 
prediction for the inclusive Higgs production cross section, cf. e.g. \cite{Anastasiou:2012hx},
Fig.~\ref{fig:HIGxs}\textcolor{black}{, which has to be considered in the ongoing 
experimental analyzes
setting exclusion limits for Higgs bosons.
The current search explores the mass range above $m_{H^0} = 114.4$~GeV \cite{Barate:2003sz}, with 
first limits being obtained in \cite{ATLAS:2012si,Chatrchyan:2012tx}.}

\restylefloat{figure}
\restylefloat{figure}
\section{Beyond the Standard Model}

\noindent
The contributions to the conference which dealt with possible extensions of the 
Standard Model concerned corrections in the MSSM \cite{BSM1,BSM2} and Kaluza-Klein models 
\cite{BSM3,BSM4,BSM5}.
There is an increasing number of studies of higher order corrections in renormalizable
extensions of the Standard Model, still awaiting experimental evidence. The LHC 
experiments have carried out numerous 
searches excluding new particles or forces below scales of $\sim 0.5-5$~TeV, depending 
on the respective model.
Key questions concerning extensions of the Standard Model are~:
\begin{itemize}
\item Which mechanism generates the masses of the present fundamental particles?
\item Do the fundamental forces unify and where? 
\item Which laws lead to the observed mass spectrum? 
\item What is the role of gravity?
\end{itemize}
\section{Summary}

\noindent
In conclusion, the field of high precision high energy physics is in good shape.
Many more precision measurements will be performed at various colliders such as
the LHC, lower energy facilities, at JPARC, and at planned facilities like a $B$-factory, 
the EIC, and ILC. The calculational tools do vastly evolve. The automation of NLO
calculations has proceeded very far and includes many important processes already.
Highly efficient numerical methods are available. Methods based on computer algebra
advance very quickly yielding deeper insight into the analytic structure of quantum
field theories. More and more new and longer known mathematical technologies contribute
to significant progress in our field. The NLO calculations reached 7-point functions.
Resummations are needed in many places to bridge between the perturbative and non-perturbative 
regions. At NNLO the calculations reached $2 \rightarrow 2(3)$ scattering processes 
including 
masses. 4-loop QCD corrections started and more are to come, and first 5-loop results are 
available. Renormalizable QFTs start to request Tbyte CPUs to solve problems analytically.
Sophisticated integrations turn more and more into algebraic problems. Many of the current 
developments were driven by high precision measurements. Both the precision reached in experiment
paired with highly accurate theoretical results lead to a very deep understanding of the 
structures currently being probed at shortest distances, resp. highest energies. We all enjoy 
to contribute to and to witness these fascinating and ground-breaking computations in one of the  most 
fundamental fields of science and urgently await to see experimental deviations of these 
predictions, which are due to new physics. 

\vspace*{2mm}
\noindent
{\bf Acknowledgment.}\\

\vspace*{-2mm}
\noindent
I would like to cordially thank the organizers of RADCOR 2011 on behalf of all 
participants for their excellent work and all their effort to make this conference
possible. I thank A. This work has been supported in part by DFG Sonderforschungsbereich 
Transregio 9, Computergest\"utzte Theoretische Teilchenphysik, and EU Network {\sf LHCPHENOnet} 
PITN-GA-2010-264564.


\end{document}